\documentclass[twocolumn]{aastex63}
\usepackage[version=4]{mhchem}

\received{\today}
\revised{\today}
\accepted{\today}

\submitjournal{ApJ}

\shorttitle{Water UV-shielding and CO$_2$ emission}
\shortauthors{Bosman, Bergin, Calahan \& Duval}

\newcommand{\vo}{\ensuremath{01^10}}
\newcommand{\vz}{\ensuremath{00^00}}
\begin{document}

\title{Water UV-shielding in the terrestrial planet-forming zone: Implications for carbon dioxide emission}


\correspondingauthor{Arthur D. Bosman}
\email{arbos@umich.edu}

\author[0000-0012-3245-1234]{Arthur D. Bosman}
\affiliation{University of Michigan, LSA astronomy \\
1085 S University \\
Ann Arbor, MI 48109, USA}

\author[0000-0003-4179-6394]{Edwin A. Bergin}
\affiliation{University of Michigan, LSA astronomy \\
1085 S University \\
Ann Arbor, MI 48109, USA}

\author[0000-0002-0150-0125]{Jenny K. Calahan}
\affiliation{University of Michigan, LSA astronomy \\
1085 S University \\
Ann Arbor, MI 48109, USA}

\author{Sara E. Duval}
\affiliation{University of Michigan, LSA astronomy \\
1085 S University \\
Ann Arbor, MI 48109, USA}

\begin{abstract}

Carbon Dioxide is an important tracer of the chemistry and physics in the terrestrial planet forming zone. 
Using a thermo-chemical model that has been tested against the mid-infrared water emission we re-interpret the \ce{CO2} emission as observed with \textit{Spitzer}.  We find that both water UV-shielding and extra chemical heating significantly reduce the total \ce{CO2} column in the emitting layer. Water UV-shielding is the more efficient effect, reducing the \ce{CO2} column by $\sim$ 2 orders of magnitude. These lower \ce{CO2} abundances lead to \ce{CO2}-to-\ce{H2O} flux ratios that are closer to the observed values, but \ce{CO2} emission is still too bright, especially in relative terms. Invoking the depletion of elemental oxygen outside of the water mid-plane iceline more strongly impacts the \ce{CO2} emission than it does the \ce{H2O} emission, bringing the \ce{CO2}-to-\ce{H2O} emission in line with the observed values. We conclude that the \ce{CO2} emission observed with \textit{Spitzer}-IRS is coming from a thin layer in the photo-sphere of the disk, similar to the strong water lines. Below this layer, we expect \ce{CO2} not to be present except when replenished by a physical process. This would be visible in the \ce{^{13}CO2} spectrum as well as certain \ce{^{12}CO2} features that can be observed by \textit{JWST}-MIRI.

\end{abstract}


\section{Introduction} \label{sec:intro}

Carbon Dioxide, \ce{CO2}, is an important molecule in the interstellar inventory.  \ce{CO2} is a carrier of significant volatile carbon and oxygen in both interstellar \citep{Boogert2015} and cometary \citep{Mumma2011} ices. However, the lack of a permanent dipole and its abundance in the Earth's atmosphere requires infrared space missions for astronomical observations.
The chemistry of \ce{CO2} is closely linked to \ce{CO}, a common precursor to \ce{CO2} and \ce{H2O}, with which \ce{CO2} shares the \ce{OH} radical as a precursor in both gas and grain-surface chemistry \citep[e.g.][]{Smith2004, Arasa2013}. Such close chemical links to the main reservoirs of carbon and oxygen make \ce{CO2} an important tracer of the carbon and oxygen elemental abundances, especially the overall C/O ratio. As the C/O ratio is proposed to link planet composition and formation history \citep[][]{Oberg2011}, \ce{CO2} likely has an important role to play in the unraveling of this complex problem. 

Towards proto-planetary disks, \ce{CO2} has most commonly been observed in emission using \textit{Spitzer}-IRS \citep[e.g.][]{Carr2008, Pontoppidan2010}. CO$_2$ emission has been inferred to originate from the warm ($>$ 300 K) surface layers confined to within the inner few au \citep{Salyk2011, Bosman2017}. In these strongly irradiated disk surface layers the abundance of \ce{CO2} is sensitive to both the C/O ratio, through the abundance of its precursor \ce{OH} \citep[e.g.][]{Woitke2018, Anderson2021}, as well as the elemental abundance of C \citep[See Fig. 1 in ][]{Bosman2018}.


The interplay between dust growth, dynamics, and chemistry in the outer (tens of au) disk might also significantly impact the abundance of \ce{CO2} in the inner few au where excitation conditions allow for detectable rovibrational emission.
The low \ce{CO} abundances in the outer disk are possibly pointing to a sequestration of carbon into \ce{CO2} ices \citep[][]{Eistrup2016, Bosman2018CO}.  This is seen in detailed models that link chemistry to dust evolution \citep[][]{Krijt2020}. As these ices drift to inside the CO$_2$ sublimation front (or ice line), they could greatly impact the inner disk \ce{CO2} abundance \citep[e.g.][]{Bosman2017, Booth2017}.   Thus CO$_2$ might be a sensitive tracer of dust drift.

Gas dynamics in the inner few au could also impact the \ce{CO2} abundance structure. Vertical mixing of water, followed by sequestration on large grains dubbed the ``vertical cold finger effect'' has been invoked to explain the limited extent of the water surface reservoir \citep{Meijerink2009, Blevins2016, Bosman2021water} and dynamical simulations have shown it to be effective \citep{Krijt2016Water}. Such sequestration would similarly affect \ce{CO2} outside of the \ce{CO2} iceline. Furthermore between the \ce{H2O} and \ce{CO2} icelines, the sequestration of water can impact the elemental abundances of Carbon and Oxygen, creating an environment in which \ce{CO2} is inefficiently formed, and the excess oxygen lost to water ice in the mid-plane. 

The abundance of CO$_2$ is a crucial component in understanding the overall C/O ratio of gas and ices, but it also can be a probe of the overall disk chemical evolution, which is linked to the disk gas and dust physics. To fully understand these processes it requires the emission be analyzed in detail.
Unfortunately, recent modeling efforts have difficulty in simultaneously predicting the \ce{H2O} and \ce{CO2} fluxes in the mid-infrared \citep{Woitke2018, Anderson2021}. In this paper, we used an updated set of thermo-chemical models that has been tested against the infrared water observations \citep{Bosman2022water} to study the \ce{CO2} abundance structure and emission from the inner disk regions. Crucially, our model includes the effects of water UV-shielding.  This must be present given the large water columns \citep[][]{Bethell2009} leading to UV absorption at higher surface layers than would be set by the dust optical surface to UV photons; this can strongly alter the chemical structure of which \ce{CO2} is a key component.  

\section{Methods}

\begin{table}[]
    \centering
        \caption{Elemental abundances w.r.t H}
    \begin{tabular}{l|c}
    \hline \hline
        Element & Abundance w.r.t. H \\
    \hline
        H &  1.0 \\
        He & $7.59 \times 10^{-2}$ \\
        C & $1.35 \times 10^{-4}$ \\
        N & $2.14 \times 10^{-5}$ \\
        O& $2.88 \times 10^{-4}$ \\
        Mg& $4.17 \times 10^{-9}$ \\
        Si& $7.94 \times 10^{-8}$ \\
        S& $1.91 \times 10^{-8}$ \\
        Fe& $4.27 \times 10^{-9}$ \\
    \hline
    \end{tabular}

    \label{tab:Chem_elem_abu}
\end{table}

We use the DALI models from \citet{Bosman2022water}. These models include modification from standard DALI \citep{Bruderer2012, Bruderer2013} to better represent the inner disk regions. This includes: more efficient \ce{H2} formation at high temperature, more efficient heating following photo-dissocation \citep[following][]{Glassgold2015}, and water UV-shielding \citep[][]{Bethell2009}. The models assume the AS 209 spectrum from \citet{Zhang2021MAPS} as input. Most of the UV in this spectrum is in Lyman-$\alpha$ \citep{Herczeg2004}. Four different physical structures are computed, two large grain fractions (99\% and 99.9\%) and two disk scale heights (h/R = 0.08 and 0.16 at R = 45 au). Model setup details are in \citep{Bosman2022water}. The elemental abundances used in the chemistry are given in Table~\ref{tab:Chem_elem_abu}. Abundances are based on \citet{Jonkheid2006}, with reduced Mg, Si, S and Fe. As the chemical time-scale in the region of interest are short, we solve for statistical equilibrium  \citep{Anderson2021}. 

From the temperature and \ce{CO2} abundance structure, we calculate the \ce{CO2} emission spectrum using the molecular data as collected and computed in \citep{Bosman2017}. The molecular excitation is calculated explicitly by balancing excitation and de-excitation from collisions and photon absorption and emission. Energy levels, line positions, and line strengths are from the HITRAN database \citep{Rothman2013}. Throughout this paper, we use the \ce{CO2} level notation as used in \citep{Bosman2017}. When talking about the main \ce{CO2} feature around 15$\mu$m we denote this as the $01^10-00^00$ $Q$-branch, but it is implied that this includes a smaller contribution of the more highly excited $02^20$--$01^10$,  $03^30$--$02^20$ , etc. $Q$-branches.   

For comparison with slab excitation models we use the model from \citet{Banzatti2012}, using representative slab model parameters from \citet{Salyk2011}. For water we use a column of $3\times 10^{18}$ cm$^{-2}$ and a excitation temperature of 500 K, as done in \citep{Bosman2022water}, while for \ce{CO2} we use a column of $3\times 10^{15}$ cm$^{-2}$ and a excitation temperature of 700 K. These effectively match the typical \ce{H2O} and \ce{CO2} spectra and are excellent proxies for the observed emission spectrum.

\section{Results}

\begin{figure}
    \centering

    \includegraphics[width = \hsize]{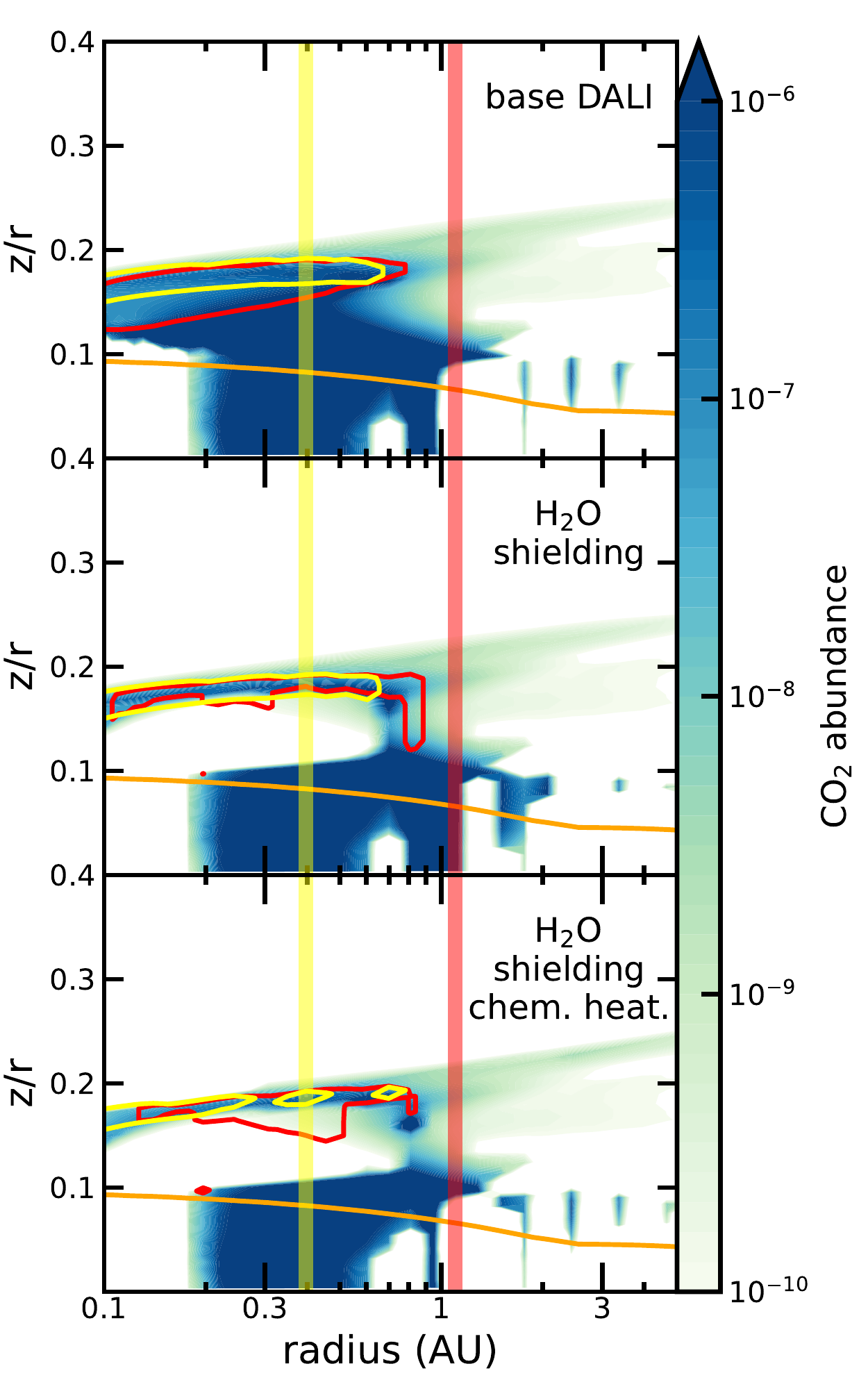}

    \caption{\ce{CO2} abundance for the thin model with a large grain fraction of 99.9\% for variation in the thermo-chemical model. Red contour shows the region from which 90\% of the \ce{CO2} \vo{}--\vz{} $R(20)$ emission line originates, the yellow contour shows the region from which 90\% of the \ce{H2O} $11_{3,9}$--$10_{0,10}$ line flux at $17.2 \mu$m originates, the orange line shows the continuum $\tau_{\mathrm{15}\mu\mathrm{m}} = 1$ surface and vertical yellow and red bands show the locations of the \ce{H2O} (0.4 au ) and \ce{CO2} (1.2 au) mid-plane icelines, respectively.  }
    \label{fig:CO2abundance}
\end{figure}

\begin{figure}
    \centering
    \includegraphics[width = \hsize]{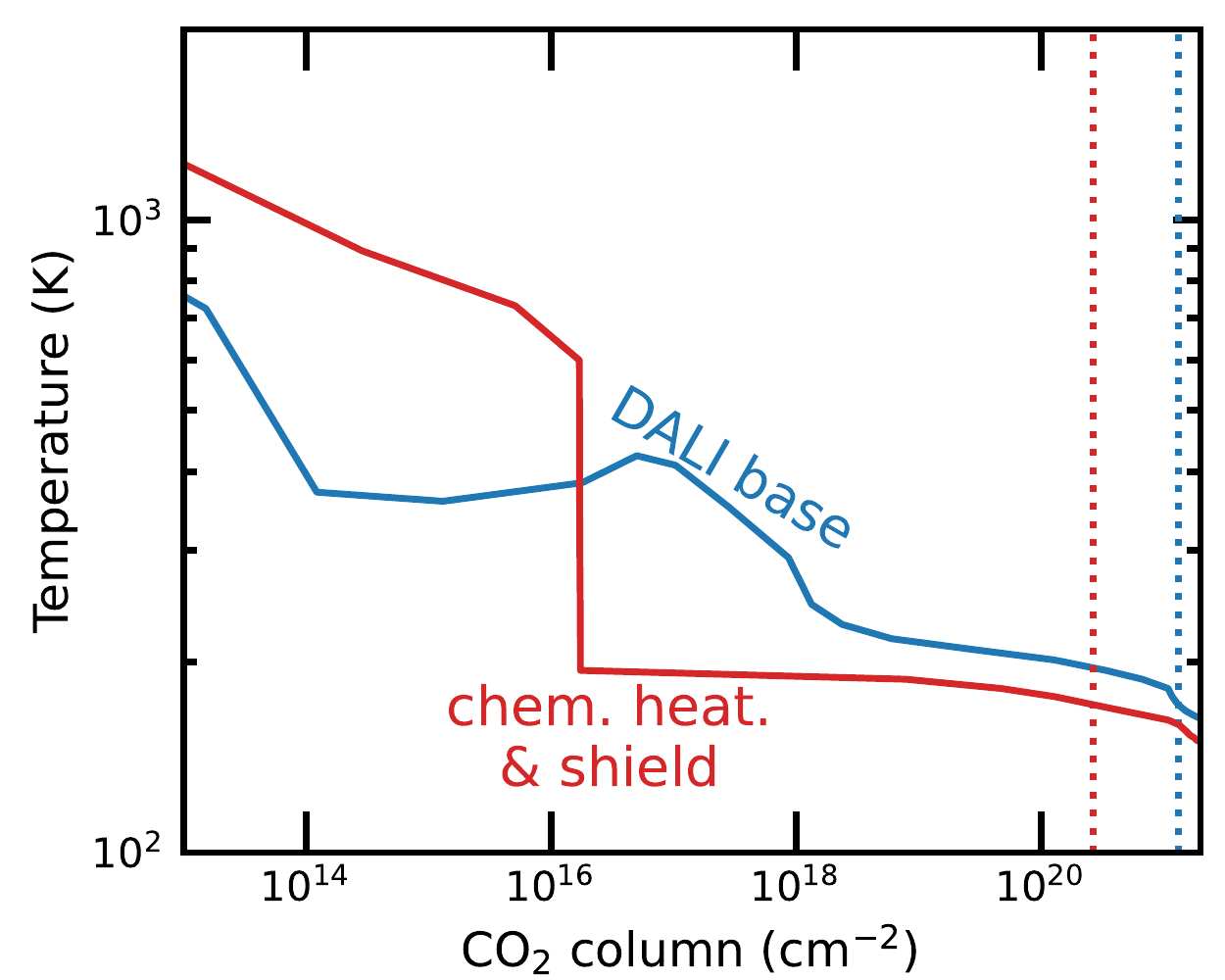}
    \caption{Gas temperature as function of the cumulative vertical \ce{CO2} column at the location of the mid-plane \ce{H2O} iceline (0.4 au). The \ce{CO2} column is a proxy for the location in the disk, higher column being deeper into the disk. In the base DALI model, \ce{CO2} is present at a significant abundance over most of the vertical extent of the molecular layer and so the \ce{CO2} temperature changes continuously with increasing \ce{CO2} column or increasing depth. In the heating and shielding model, after reaching a column of about $10^{16}$ cm$^{-2}$ at a temperature of 600 K, the \ce{CO2} abundance drops significantly. Only very deep into the disk when the temperature has dropped down to 200 K, does the \ce{CO2} abundance reach appreciable levels and does the vertical \ce{CO2} column increase again with depth. This causes the strong jump in the temperature profile.  }
    \label{fig:Vert_T_column}
\end{figure}

\begin{figure*}
    \centering
    \includegraphics[width = \hsize]{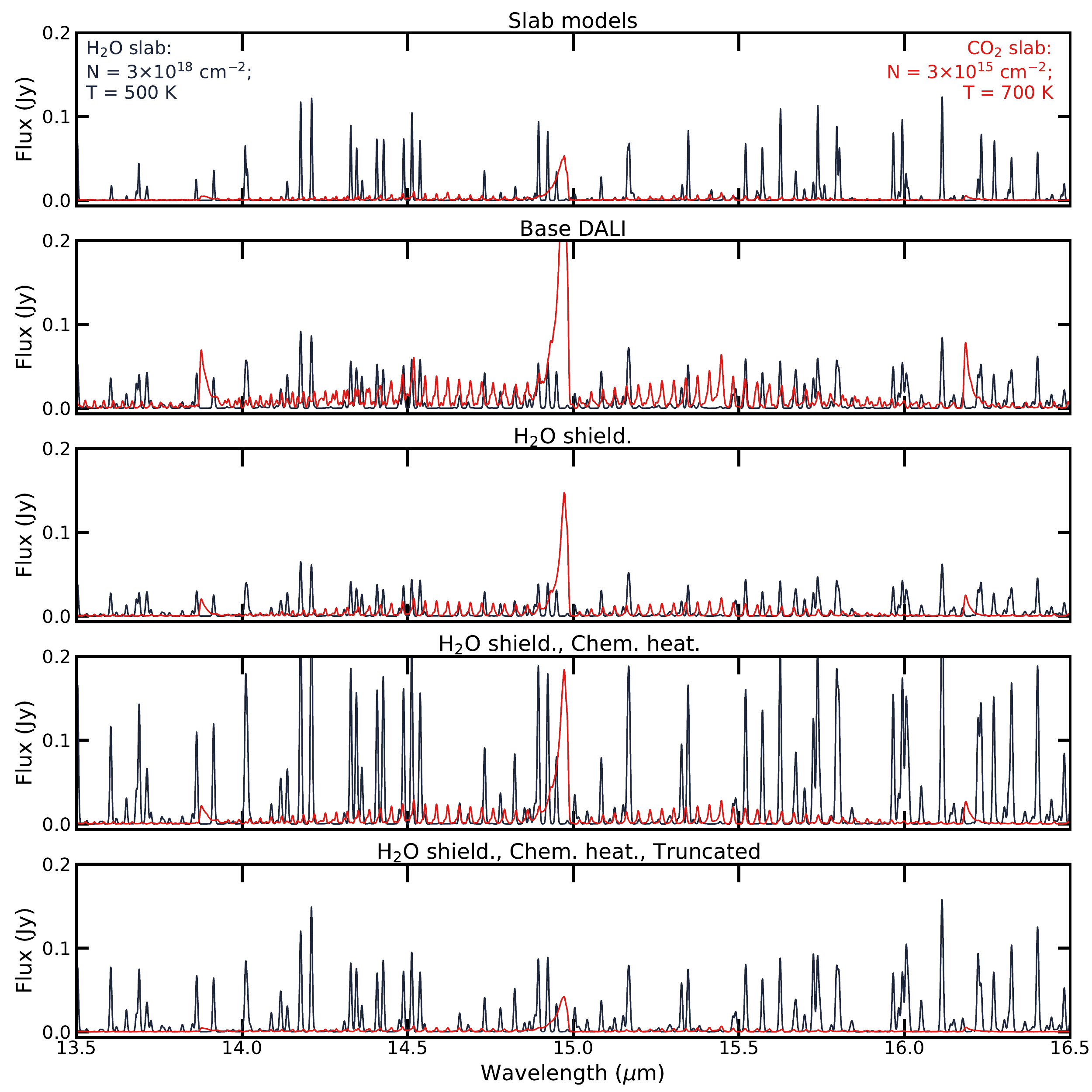}

    \caption{\ce{H2O} and \ce{CO2} spectra convolved to a R of 3000 around the \ce{CO2} \vo--\vz{} transition. Top panels shows slab model spectra using parameters that reproduce the observational data \citep{Salyk2011}. The four panels below show the DALI output spectra for the thin disk structure with 99.9\% large, settled grains. The second panel shows the base DALI model, the third base DALI with water UV-shielding, the fourth base DALI with water UV-shielding and extra chemical heating and the fifth panel shows the same model as the fourth panel, with now assuming emission can only originate from the region within the water mid-plane iceline. The slab models clearly predict a \ce{CO2} feature at 15 micron that is weaker than the surrounding waterlines, while most DALI models show a brighter \ce{CO2} feature compared to the water lines. Only when combining water UV-shielding, extra chemical heating and a restricted radial emitting region does the thermo-chemical flux ratio come close to the observed flux ratio as exemplified by the slab models. }
    \label{fig:Spectra_thinE5}
\end{figure*}

\subsection{\ce{CO2} abundance}

Figure~\ref{fig:CO2abundance} shows the \ce{CO2} abundance structure for the flat model with a large grain fraction 99.9\%. In this baseline model, the \ce{CO2} abundance is roughly distributed in two reservoirs.  On the disk surface there is the warm ($>$ 500~K), \ce{CO2} layer with an abundance between $10^{-8}$-- $10^{-6}$ spanning the entire 0.1--1~au region, with a low abundance tail to larger radii. Below this is a high abundance $>10^{-6}$, cooler (100-500 K) \ce{CO2} region that reaches all the way to the mid-plane, and out to the \ce{CO2} iceline.  There is some additional structure  as CO$_2$ vapor is absent in the inner $\sim$ 0.2 au and in a region around 0.7 au near the disk mid-plane, the former is caused by carbon sequestration in hydrocarbons, while the latter is caused by the efficient formation of water ice. The high abundance, cool \ce{CO2} reservoir extends to high in the disk photo-sphere (z/r $> 0.1$) and contributes to line formation. This \ce{CO2} morphology is identical for the other physical structures. 

When \ce{H2O} UV-shielding is included, both reservoirs are significantly impacted (see second panel of Fig.~\ref{fig:CO2abundance}). The warm surface layer becomes thinner and the region of high CO$_2$ abundance ($\sim$10$^{-6}$) is restricted to below z/r $\le$ 0.1.
In this model, \ce{CO2} vapor returns to the high abundance value around the $\tau_{\mathrm{15}\mu\mathrm{m}} = 1$ layer. The decrease of the \ce{CO2} abundance is driven by the slower photo-dissociation of \ce{H2O}, leading to lower \ce{OH} production rates, which is critical for the production of \ce{CO2}. Furthermore the \ce{CO2} dissociation rate is not decreased by the same amounts as the \ce{H2O} dissociation rate. Unfortunately the OH emission from disks is dominated by \ce{OH} that is above the point that \ce{H2O} can start to block the UV radiation, as such OH cannot be used as tracer of \ce{H2O} UV-shielding (see Appendix~\ref{app:OH}).  

Increasing the chemical heating after photo-dissociation only impacts the abundance in the warm surface layer. The increased temperatures increase the efficiency of the \ce{OH + H2 -> H2O + H} reaction which has a high reaction barrier \citep[1740 K;][]{Baulch1992}, leaving less \ce{OH} for the \ce{OH + CO -> CO2 + H} reaction, changing the \ce{H2O} to \ce{CO2} abundance ratio \citep[see, e.g., Fig.~1 of][]{Bosman2018}. 

Figure~\ref{fig:CO2abundance} also shows the emitting area of the \vo{}--\vz{} $ R(20)$ line as an irregular red box. This line has an upper level energy of $\sim$1250 K of which $\sim$250 K is in rotational energy. This is relatively low in  comparison to the nearby water lines, many of which have upper level energies of $>$2500 K. The emitting region of this line clearly traces the warm surface layer of the disk and extends over the  full radial extent in gas where the \ce{CO2} vapor abundance lies between 10$^{-9}$ to few $\times 10^{-7}$. Lines with lower rotational excitation will emit from a slightly larger radial region. As can be seen in Fig~\ref{fig:CO2abundance} the \ce{CO2} emitting area extends to layers exterior to the water mid-plane iceline and, in almost all cases, the \ce{CO2} emitting area is larger than that for the water vapor lines.

Figure~\ref{fig:Vert_T_column} shows the gas temperature as function of the vertical, top-down \ce{CO2} column at the radius of the water iceline. Here, for ease in comparison we show only two models, both with large grain fractions of 99.9\%, one is the base DALI model and the other with extra chemical heating and water UV-shielding.  These two models show sharply different structure.
In the non-shielding model, \ce{CO2} is abundant throughout the full vertical extent of the disk, so the temperature decreases relatively smoothly with increasing \ce{CO2} column. In the water UV-shielding model, \ce{CO2} is absent in the intermediate layers. As a result the temperature drops sharply when the column reaches around $10^{16}$ cm$^{-2}$. The low \ce{CO2} abundance in the intermediate layers is caused by the UV shielding of water. The self-shielding of water stops the formation of OH, inhibiting the formation of \ce{CO2}. As \ce{CO2} can dissociate at longer wavelengths compared to \ce{H2O}, dissociation of \ce{CO2} is still possible, together this strongly lowers the \ce{CO2} abundance. In the deeper layers, all UV is blocked by the small dust. In these layers, that have very high density the statistical-equilibrium that is calculated tends towards the chemical-equilibrium. 

This abundance and temperature structure strongly constrains where the emission in the water UV-shielded model can originate from. Only a column of $\sim 10^{16}$ cm$^{-2}$ can contribute to the emission. \ce{CO2} deeper into the disk exists at much lower temperatures, as well as at temperatures close to the $\tau_{\mathrm{15}\mu\mathrm{m}} = 1$ surface ($\sim$200 K). Contribution for columns $>10^{16}$ cm$^{-2}$ should thus be negligible. Contrast this with the models without shielding, where the temperature is $\sim$ 300 K at a column of 10$^{18}$ so two orders of magnitude higher columns can contribute to the emission in this model.

\subsection{\ce{CO2} spectra}

Figure~\ref{fig:Spectra_thinE5} shows the \ce{H2O} and \ce{CO2} spectra around the 15 $\mu$m \ce{CO2} vibrational band. The baseline DALI models predicts relatively stronger \ce{CO2} emission when compared to the surrounding \ce{H2O} line forest. The \ce{CO2} \vo--\vz{} Q-branch is a factor few brighter than the surrounding water lines and the individual \ce{CO2} \vo--\vz{} $P$ and $R$ branch lines are on par with the water lines. Both of these characteristics are inconsistent with Spitzer/IRS observations of typical disk systems as exemplified with the slab models in Fig.~\ref{fig:Spectra_thinE5}. 

Including water UV-shielding reduces the \ce{CO2} flux. This is a direct consequence of the thinner upper atmosphere \ce{CO2} layer. As the water emission is less strongly affected by the inclusion of water UV-shielding \citep{Bosman2022water}, the ratio between water and \ce{CO2} lines gets closer to the ratio in the slab models, but (in a relative sense) CO$_2$ emission is still too bright. 

Including extra chemical heating, and thus increasing the gas temperature correspondingly leads to higher \ce{CO2} infrared flux. However, as the increased temperature lowers the \ce{CO2} abundance, the flux ratio between water lines and the main \ce{CO2} feature decreases. As a result, the model with both water UV-shielding as well as chemical heating comes closest to the observed \ce{H2O}-to-\ce{CO2} ratios. \ce{CO2} in this model remains too bright when compared to the surrounding water emission. However, as noted earlier, the \ce{CO2} emitting region extends further than the \ce{H2O} emitting region. Additional spectra from different physical structures find similar results (See App.~\ref{app:struct_variation}). 

Figure~\ref{fig:Spectra_thinE5} also shows the \ce{H2O} and \ce{CO2} spectra in the case where we assume that the emission of both is contained radially to within the water mid-plane iceline (see fig.~\ref{fig:CO2abundance}), perhaps due to the vertical cold finger effect \citep[][]{Meijerink2009}.   This cuts both the \ce{H2O} and \ce{CO2} flux, but in general the \ce{CO2} lines are more strongly impacted than the \ce{H2O} lines. In the case of the model with shielding, extra chemical heating and a truncated emitting region the model \ce{H2O} and \ce{CO2} spectra line up nicely with the slab model spectra that are representative for the {\it Spitzer}-IRS observations.

\section{Discussion}

\subsection{\ce{CO2} column}

The thermo-chemical model with water UV-shielding and chemical heating consistently predict a \ce{CO2} column in the surface layers that is around $10^{16}$ cm$^{-2}$. This is orders of magnitude lower than the $10^{18}$ cm$^{-2}$ columns that follow from the models that do not include water UV-shielding.  However, it is on the high end of the observed data where columns span the $3\times 10^{14}$--$3\times10^{16}$ cm$^{-2}$, with additional non-detection implying even lower \ce{CO2} columns. As such our models are not capturing the full variation in the observed population. They are however representative of a significant part of the full population. 

The gas temperatures in our model at the outer edge of the \ce{CO2} emission is colder ($<$500 K) than the inferred temperatures from the \ce{CO2} \textit{Spitzer}-IRS observations  \citep[$\sim$700 K;][]{Salyk2011}. This could imply that the radial temperature profile in the disk is impacting the shape of the \vo--\vz{} $Q$-branch forcing the slab-model fit to a higher temperature, but lower column solution. This could also explain why the slab models consistently find a higher gas-temperature for the \ce{CO2} than for the \ce{H2O} \citep[][]{Salyk2011}, whereas in the models they both emit from the same gas. With individual $P$ and $R$-branch lines of \ce{CO2} and better isolated lines of \ce{H2O} in the JWST-MIRI spectra, it should be possible to get a better handle on the gas-temperature and column densities of both these species. 

There are chemical pathways that could lead to lower \ce{CO2} abundances. Increasing the gas-temperature changes the balance between \ce{H2O} and \ce{CO2} formation, which both rely on OH; as a result a higher temperature lowers the \ce{CO2} abundance. A temperature increase requires either a larger total UV flux or mechanical heating, as currently most of the energy in the UV photons is already converted into heat in the models \citep{Bosman2022water}.

Another possible way to lower the \ce{CO2} abundance (and decrease emission) is to increase its destruction rate. Two and three-body gas-phase reactions that destroy \ce{CO2} are few and inefficient \citep[see][]{Bosman2018}, whereas reactions due to an ionization source (cosmic-rays and X-rays) are already included. This leaves the dissociation of \ce{CO2} due to UV, which already is the main destruction pathway in the surface layers where \ce{CO2} emission is being produced. The current stellar spectrum contains a significant amount of Ly-$\alpha$.  The photodissociation cross-section of \ce{CO2} is modest near Ly-$\alpha$ and \ce{CO2} is not efficiently dissociated in our model \citep[][]{Huestis2010, Archer2013, Heays2017}. However, these cross-sections are from measurements  at room temperatures. At elevated temperatures, the UV cross section at Ly-$\alpha$ increases by up to two orders of magnitude \citep{Venot2018}. Implementing these high temperature cross sections, however, had little effect on the \ce{CO2} distribution or spectra. In these models  the \ce{CO2} abundance decreased by 10--20\% in the disk surface layers. As such, we deem that our chemical model is robust in its treatment of \ce{CO2} destruction pathways. 

\subsection{Tracing the \ce{CO2} abundance structure}

To properly interpret the \ce{CO2} mid-infrared flux, and derive the elemental composition of the gas and explore disk physical processes, it is important to understand distribution of \ce{CO2} throughout the disk. The models with water UV-shielding imply that \ce{CO2} will only be present in a thin surface layer and will emit at a similar temperature as water vapor. However, our models predict that water UV-shielding has a strong influence on the CO$_2$ vertical abundance distribution (Fig.~\ref{fig:CO2abundance}) which has significant impact on the column distribution within different thermal layers (Fig.~\ref{fig:Vert_T_column}).

\begin{figure*}
    \centering

    \includegraphics[width = \hsize]{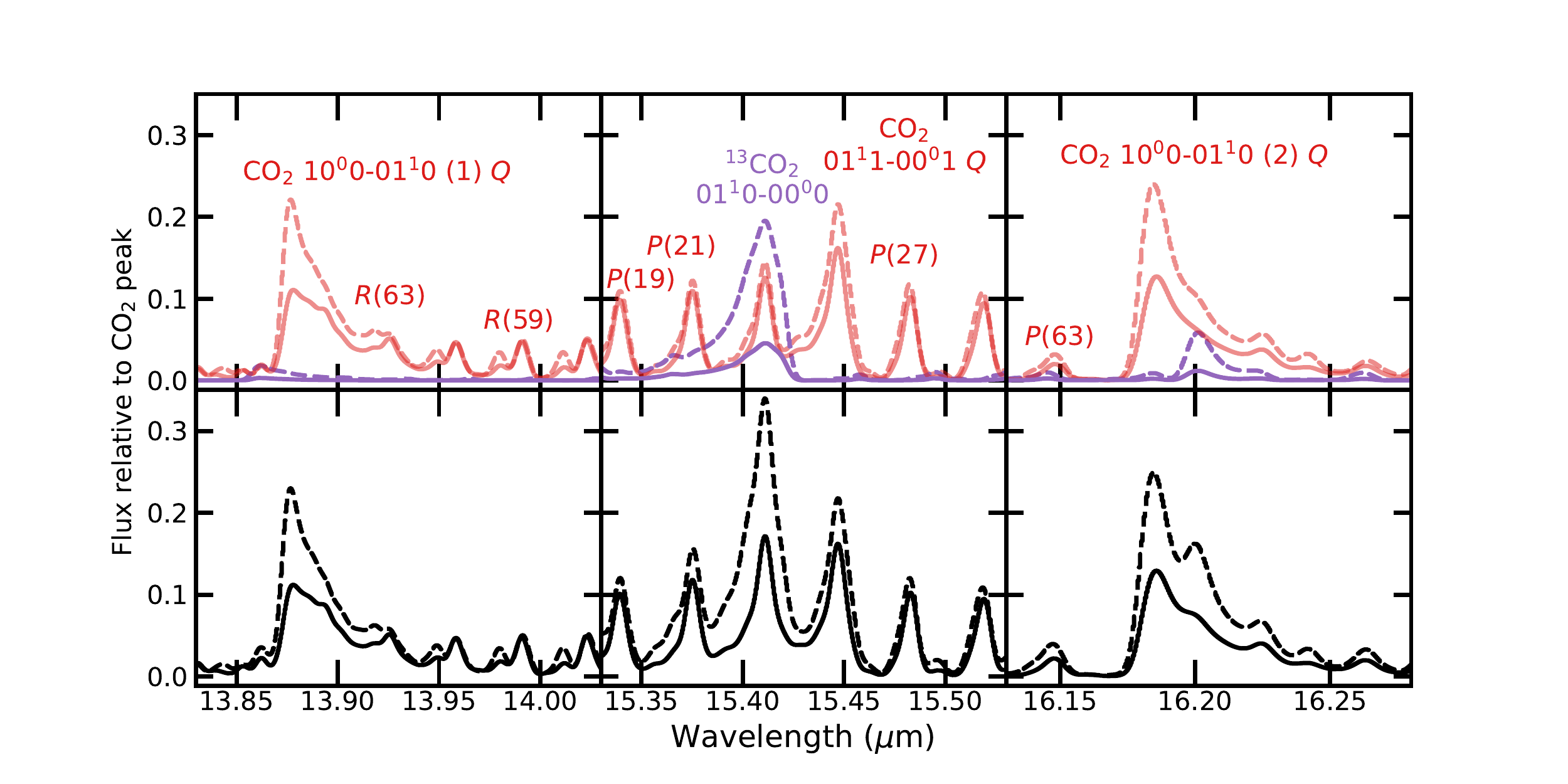}

    \caption{Normalized spectra convolved to a R=3000 for the chemical heating model with (solid) and without (dashed) water UV-shielding assuming a thin structure with a surface layer gas-to-dust ratio of  $10^5$. \ce{CO2} is assumed to only be present within the water mid-plane radius for all these models. The top panel shows \ce{^{12}CO2} and \ce{^13CO2} in red and purple respectively. The bottom panel shows the combined spectra. Spectra have been normalize to the \ce{^{12}CO2} \vo--\vz{} $Q$-branch peak and are zoomed in on three regions with $Q$-branches of interest. While the individual \vo--\vz{} $P$ and $R$-branch lines show little differences in normalized flux, the various $Q$-branch features are impacted by the  different abundance structures. Ratios between these can thus be used to derived the deeper abundance structure.  }
    \label{fig:Spect_zoom}
\end{figure*}

With higher spectral resolution when compared to Spitzer/IRS, JWST (NIRSPEC and MIRI) offer an opportunity to observe \ce{CO2} vapor with MIRI being able to isolate individual P and R branch lines around 15 $\mu$m and potentially detect \ce{^{13}CO2} \citep{Bosman2017}. In Fig.~\ref{fig:Spect_zoom} we predict \ce{^{12}CO2} and \ce{^{13}CO2} features that can help elucidate the vertical abundance structure of \ce{CO2}. In the case without water UV-shielding, there is a significantly higher \ce{CO2} abundance in the deeper layers below the nominal emitting surface region (see, Fig.~\ref{fig:CO2abundance}). In this case, the relative strength of the two $10^00-01^10$ Q-branch features at 13.90 and 16.20 $\mu$m is high, 20\% of the main $Q$-branch, while in the model without water UV-shielding, these features are about a factor 2 weaker. \ce{^{13}CO2} emission  is also sensitive to these differences; this is in line with previous predictions \citep{Bosman2017}. Conversely, the models predict a nearly constant relative strength of the \vo--\vz{} $P(19)$--$P(27)$ and $R(19$)--$R(27)$ lines, around 10\% w.r.t. the main \ce{CO2} $Q-$branch. These lines can thus be used as a yard stick. If the peak of $Q-$branch features at 13.90 and 16.20 $\mu$m is higher than the than the \vo--\vz{} rotational lines, this indicates deep, warm \ce{CO2}. It should be noted that many of the \ce{CO2} emission features have nearby \ce{H2O} lines, that will be blended in observations. Simultaneous fitting of the water spectrum is thus required.

\subsection{Radial extent of the emitting layer}

In the chemical models the radial extent of the emitting layer is naturally restricted by the presence of \ce{OH}. Formation of \ce{OH} is only efficient at high enough temperatures that the reaction barrier of the \ce{O + H2} reaction can be over come. This effectively sets the maximal \ce{H2O} emitting area. As chemically \ce{CO2} is linked to \ce{H2O} through the presence of \ce{OH}, \ce{CO2} is thus bound to a similar region from which emission can arise as the \ce{H2O} emitting region. 

\ce{CO2}, however, due to the lower upper level energies of the main transitions around 15 $\mu$m, emits more strongly from colder gas a large radii when compared to water transitions accessible to JWST. This is compounded by the higher \ce{CO2} abundance in colder gas. Radially restricting the emission to warmer gas impacts the \ce{CO2} emission more strongly than the \ce{H2O} emission. This can be seen in Fig.~\ref{fig:Spectra_thinE5}. As the radially restricted emission reproduces the \ce{CO2}-to-\ce{H2O} flux ratios from the observations. This implies that a the emitting areas are actually restricted more tightly than the chemistry implies and that a physical process like the ``cold finger effect'' is active \citep[e.g.][]{Meijerink2009, Bosman2021water}. This process impacts the \ce{H2O} directly by mixing it downward and locking up the water in the midplane ices \citep[see, e.g.][]{Krijt2016Water}.  This, in turn, impacts the chemical equilibrium of \ce{CO2} due to  the lower availability of oxygen, critical for the formation of \ce{CO2}. As \ce{CO2} is constantly destroyed in the surface layers by UV, any available gaseous \ce{CO2} between the \ce{H2O} and \ce{CO2} mid-plane icelines will constantly be reprocessed into \ce{H2O} which can then be trapped in the mid-plane ice. We note that trapping \ce{CO2} in the \ce{H2O} ice would still create a \ce{CO2} rich atmosphere outside of the \ce{H2O} iceline due to back diffusion of the \ce{CO2} gas \citep{Bosman2018}.

This could be relatively easily tested by comparing velocity resolved line profiles of \ce{CO2} and \ce{H2O}, unfortunately \ce{CO2} is hard to do from the ground \citep[e.g.][]{Bosman2017} and currently no planned space mission that will cover this wavelength range with a medium or high spectral resolving power ($\lambda/\Delta\lambda > 10000$). As such, the emitting area from \ce{CO2} will have to be extracted from analysis of the velocity unresolved spectra. Comparison with the 4.3 $\mu$m band which can be observed with JWST-NIRSpec could be helpful here as these lines originate from a radially more constrained region and are thus less affected by the abundance of \ce{CO2} outside of the \ce{H2O} midplane iceline (see Appendix~\ref{app:CO2_43}). The feature strength ratio between the 4.3 $\mu$m band and the 15$\mu$m band could thus be a measure for the radial extent of the total \ce{CO2} emitting region. 

\section{Conclusions}

We study the impact of water UV-shielding and chemical heating on the predicted \ce{CO2} emission arising from within the terrestrial planet-forming zones of gas-rich disks around 15 $\mu$m using the thermo-chemical model DALI. Our base model finds a similar result as found in previously published models \citep{Woitke2018, Anderson2021} in that the \ce{CO2} is too bright relative to the \ce{H2O} emission. 

We find that the water UV-shielding has a strong impact on the \ce{CO2} abundance structure. With UV-shielding included, disk photospheric \ce{CO2} is constrained to a thin layer close to the H/\ce{H2} transition. This yields a \ce{CO2} column of a few times $10^{16}$ cm$^{-2}$, two orders less than the \ce{CO2} column prediction without water UV-shielding, $\sim${}$10^{18}$ cm$^{-2}$. The lower \ce{CO2} column is in line with the observationally derived column. This suggests that water-UV shielding is present in disk systems, which is consistent with simple conclusions based on the inferred water vapor column \citep[][]{Bethell2009}.
This large change in column has a strong effect on the \ce{CO2} emission, both in absolute terms, but particularly in the flux ratio between the main isotopologue \vo--\vz{} $Q$-branch on the one hand and the two $10^00-01^10$ Q-branch features at 13.90 and 16.20 $\mu$m or the main  \ce{^{13}CO2} $Q$-branch on the other hand.

We find the best agreement between thermo-chemical model spectra prediction and observations, if we invoke a radial restriction of the emitting area of both \ce{CO2} and \ce{H2O} on top of the inclusion of water UV-shielding and extra chemical heating. The radial restriction increases the average temperature of the emitting gas, which appears to be required by observations. This has an impact on the chemistry, lowering the \ce{CO2}-to-\ce{H2O} abundance, as well as favoring the \ce{H2O} lines relative to the \ce{CO2} lines as the \ce{H2O} lines have higher upper level energies than the main \ce{CO2} lines in the spectra. 

We propose that this is evidence of the ``cold finger effect'', that is the sequestration of oxygen in the form of water on the mid-plane ice outside of the mid-plane water iceline. This process would lower the oxygen abundance until the C/O ratio reaches unity, locking all oxygen in CO and preventing the formation of both \ce{H2O} and \ce{CO2}, which rely on the presence of O or OH, outside of the water mid-plane iceline, even in the disk atmosphere.

\acknowledgments 

The authors thank the referees for their constructive reports that improved the quality of the paper. 
ADB and EAB acknowledge support from NSF Grant\#1907653 and NASA grant XRP 80NSSC20K0259. 
\software{Astropy \citep{astropy2013,astropy2018}, SciPy \citep{Virtanen2020},  NumPy \citep{van2011numpy}, Matplotlib \citep{Hunter2007}, DALI \citep{Bruderer2012, Bruderer2013}}

\bibliography{Lit_list}{}
\bibliographystyle{aasjournal}

\appendix

\section{OH spectra}
\label{app:OH}
The Hydroxide radical, OH, is a critical intermediate in the \ce{CO}, \ce{CO2} and \ce{H2O} chemistry. As such it is worth looking at the effects of \ce{H2O} UV-shielding and chemical heating on the \ce{OH} emission. It should be noted that calculating \ce{OH} emission is more complex than \ce{H2O} and \ce{CO2} emission. \ce{H2O} and \ce{CO2} emission are set by collision and radiation equilibrium.  \ce{OH} emission contains a contribution from highly excited \ce{OH} that is produced in \ce{H2O} photo-dissociation \citep[e.g.][]{Carr2014, Tabone2021}. Furthermore \ce{OH} collisional rate coefficients are not publicly available at present.  Therefore, we calculate the \ce{OH} excitation in LTE. This has an unfortunate side effect that OH emission arises predominantly low density gas at radii $>$10 au, yielding unrealistically high fluxes with line-to-continuum ratios $>10$ when convolved to an R of 3000. Fixing all of these issues would be possible with a non-LTE model, and possibly a density-dependent chemical heating rate in the gas, but that is beyond the scope of this paper. As such we only consider a model with the OH abundance truncated to within the water mid-plane iceline and do an inter-model comparison only.    

\begin{figure*}
    \centering
    \includegraphics[width = \hsize]{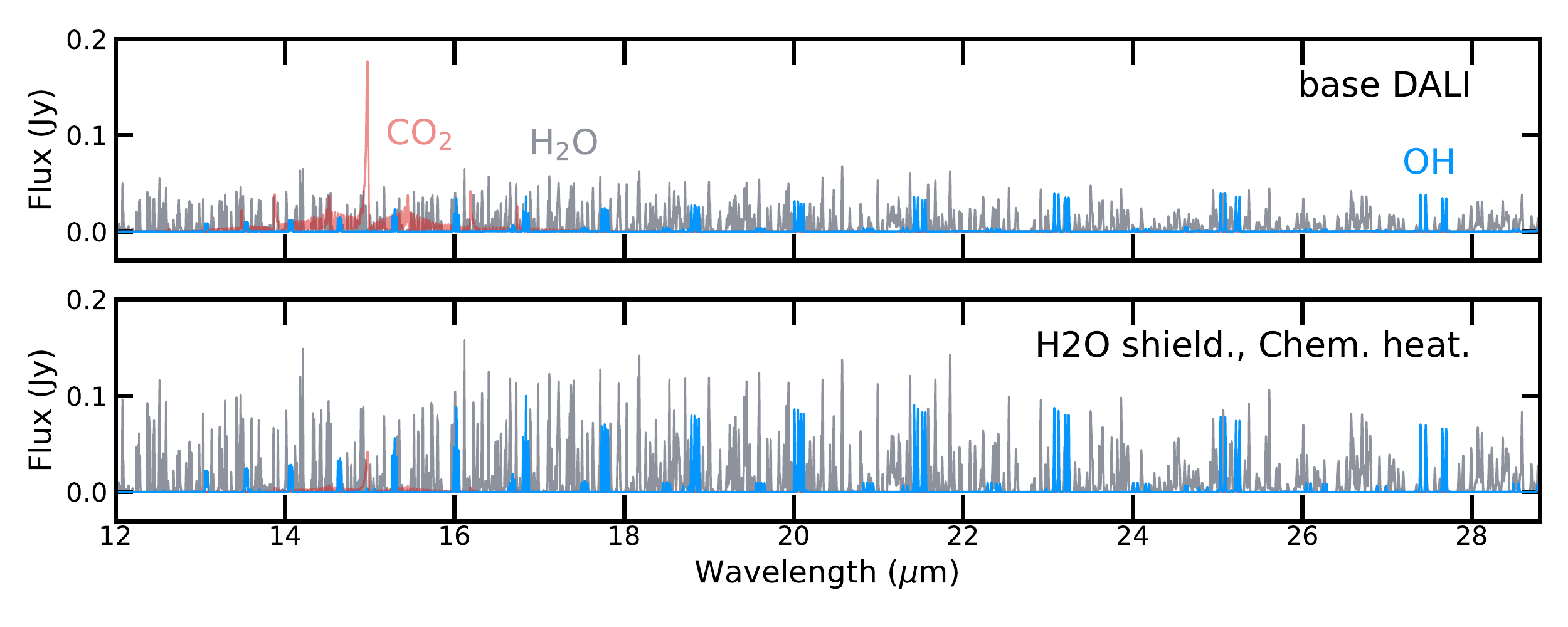}
    \caption{Spectra of \ce{OH}, \ce{CO2} and \ce{H2O} for the base and \ce{H2O} UV-shielding and chemical heating models. The emitting region has been constrained to the radial region within the water mid-plane iceline and the spectra have been convolved to R=3000. LTE excitation is assumed for OH.  } 
    \label{fig:OH_constr}
\end{figure*}

The OH LTE spectra are show in Fig.~\ref{fig:OH_constr} in comparison to the \ce{H2O} and \ce{CO2} spectra. Interestingly, the \ce{OH} to \ce{H2O} line flux seems to be similar between the base and \ce{H2O} shielding and heating model. This is due to the \ce{OH} column barely changing between the two models. The OH column is dominated by the \ce{OH} in the top layers of the atmosphere, when the density of \ce{H2} gets high enough to efficiently form \ce{H2O}, the \ce{OH} abundance drops significantly and the \ce{OH} column barely increases with increasing depth from that point onward. As such the only difference between the models is the temperature at which OH emits. The temperature change for the \ce{OH} and \ce{H2O} emitting regions are similar, as such their line ratio does not change significantly. However, with the large number of non-LTE effects that can influence the \ce{OH} emission, it definitely worth looking into \ce{OH} emission in more detail in a further study. 

\section{Model spectra for different physical structures}
\label{app:struct_variation}
\begin{figure*}
    \centering

    \includegraphics[width = \hsize]{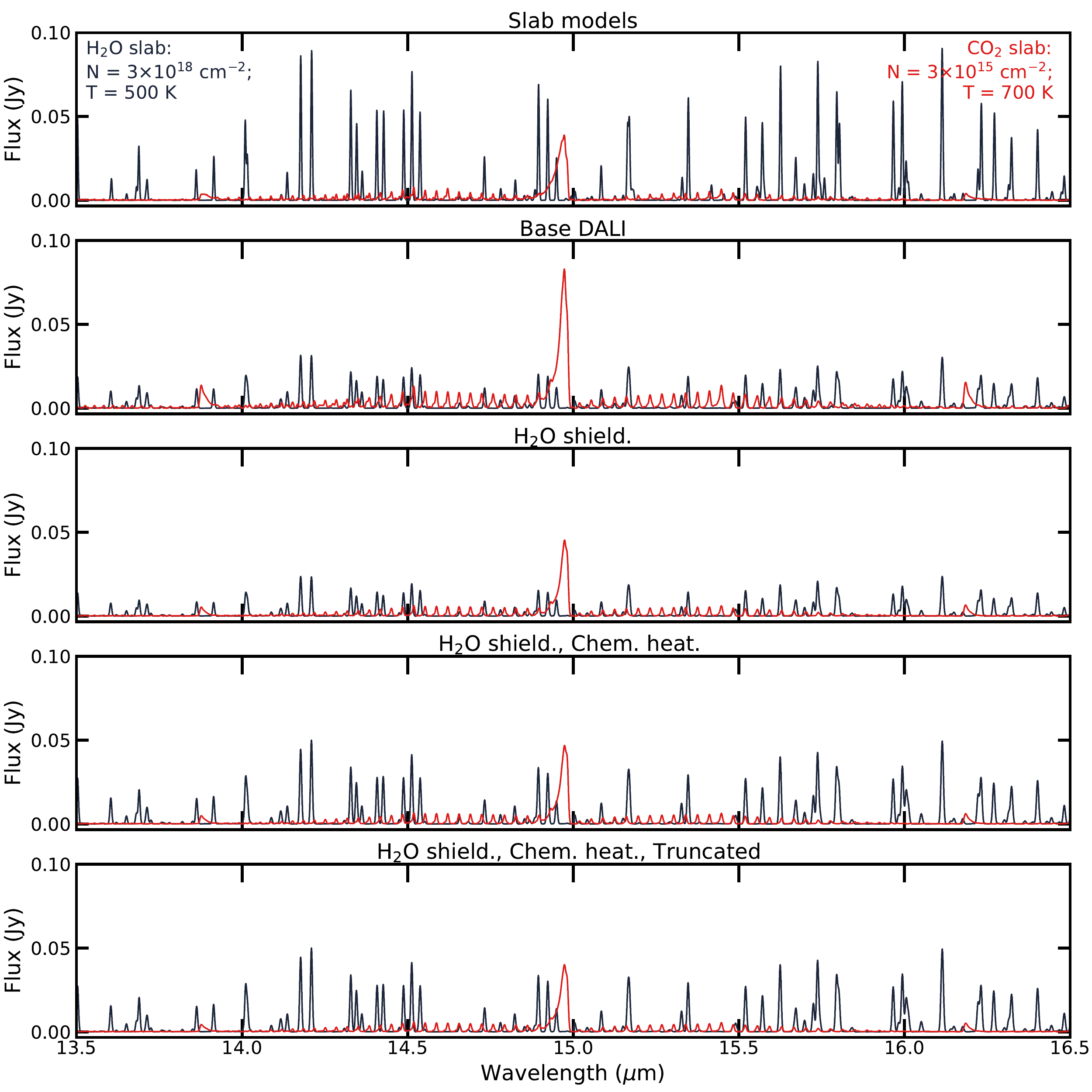}

    \caption{As Fig.~\ref{fig:Spectra_thinE5} but from the thin model with a surface layer gas-to-dust ratio of $10^4$. }
    \label{fig:Spectra_thinE4}
\end{figure*}

\begin{figure*}
    \centering

    \includegraphics[width = \hsize]{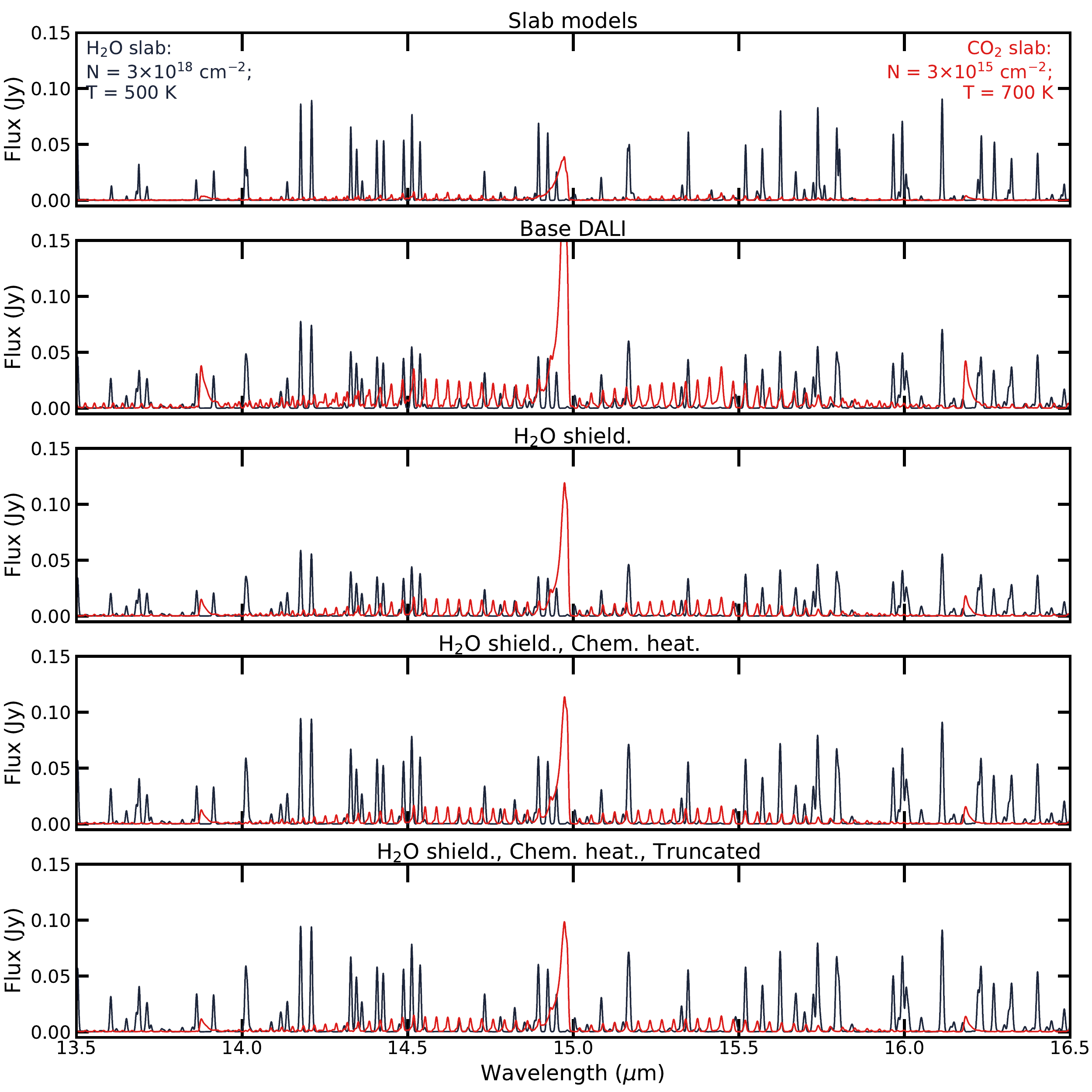}

    \caption{As Fig.~\ref{fig:Spectra_thinE5} but from the thick model with a surface layer gas-to-dust ratio of $10^4$. }
    \label{fig:Spectra_thickE4}
\end{figure*}

\begin{figure*}
    \centering

    \includegraphics[width = \hsize]{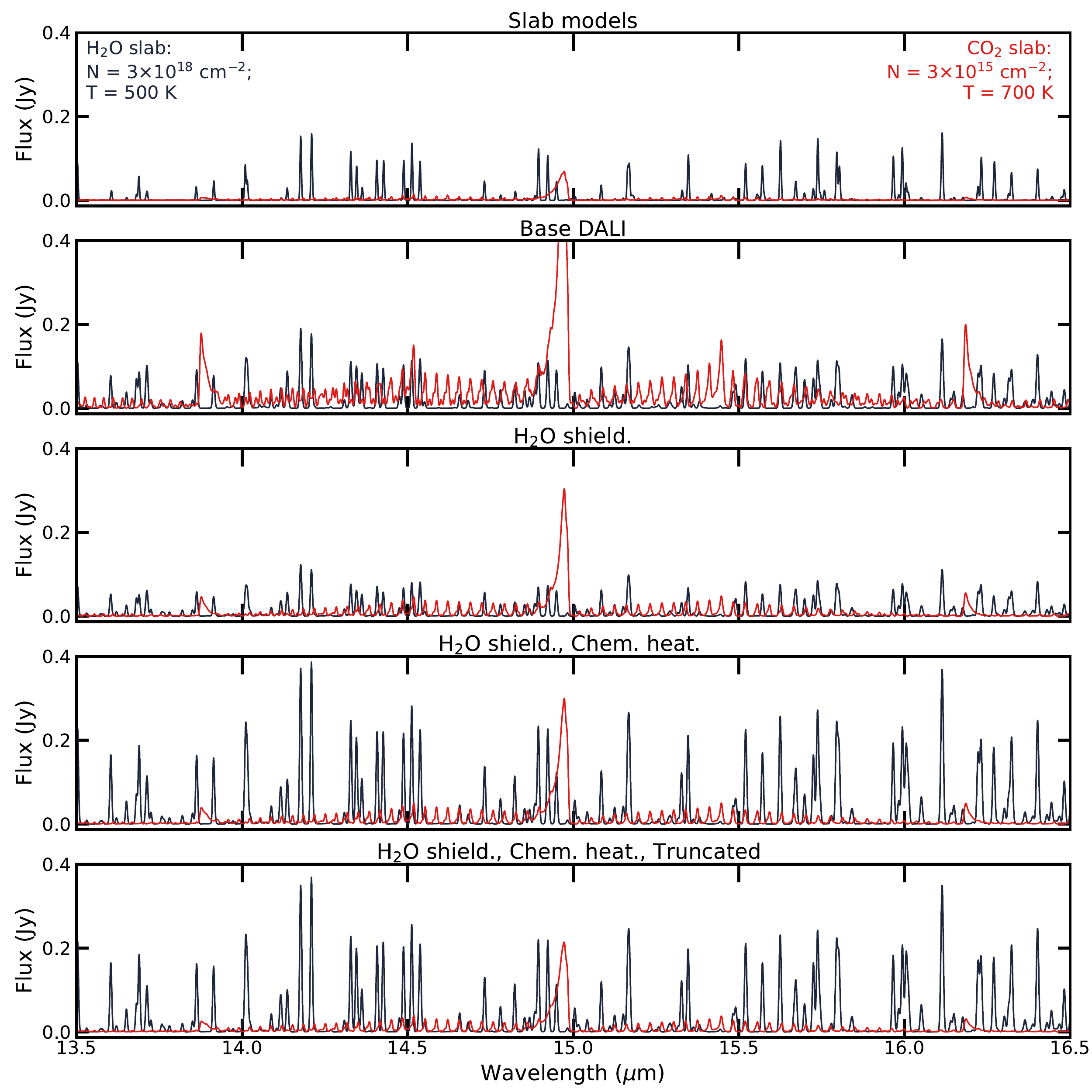}

    \caption{As Fig.~\ref{fig:Spectra_thinE5} but from the thick model with surface layer gas-to-dust ratio of $10^5$. }
    \label{fig:Spectra_thickE5}
\end{figure*}

Figures~\ref{fig:Spectra_thinE4},~\ref{fig:Spectra_thickE4}~and~\ref{fig:Spectra_thickE5} show the \ce{H2O} and \ce{CO2} spectra for different variations of the physical structure. The same general trends hold as for the spectra for the reference structure in Fig.~\ref{fig:Spectra_thinE5}. The variations show, however that both more dust, as well as a more puffed up structure lead to a strong \ce{CO2} $Q$-branch relative to the surrounding water lines. More dust leads to lower gas-temperatures in the surface layer, which promotes the \ce{CO2} production. 



\section{Predictions for \ce{CO2} at 4.3 $\mu\mathrm{m}$}
\label{app:CO2_43}
\begin{figure*}
    \centering
    \includegraphics[width = \hsize]{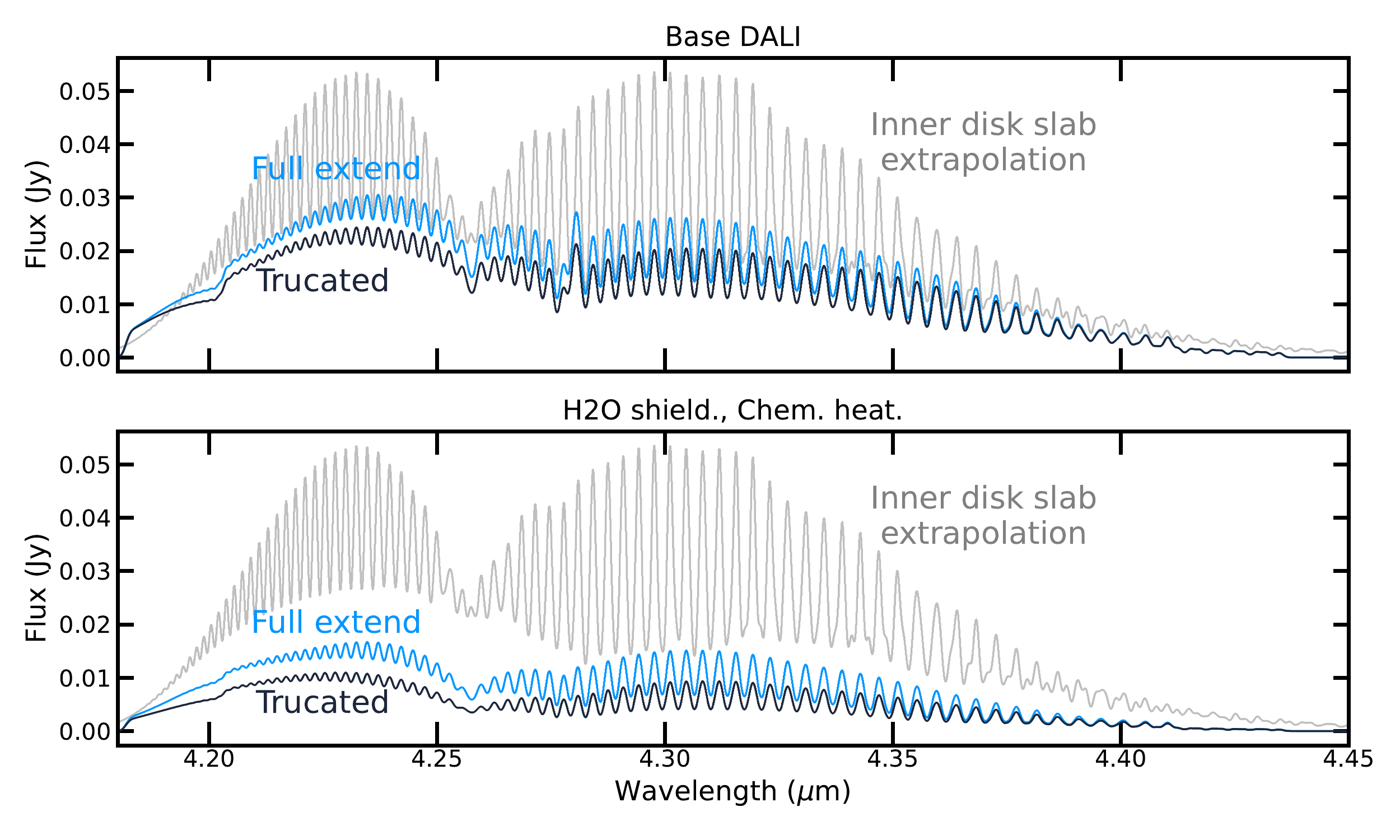}
    \caption{\ce{CO2} spectra around 4.3$\mu$m convolved to R=3000. Most flux is produced by the 00$^0$1(1)--00$^0$0(1) band, but there also is a contribution of the 01$^1$1(1)--01$^1$0(1) band. The top panel shows the spectra including the full extent of the \ce{CO2} emission for base DALI (light blue) and our complete model with \ce{H2O} UV-shielding and extra chemical heating. The bottom panel shows the same models, but with the \ce{CO2} emission constrained to within the \ce{H2O} mid-plane iceline. The light grey spectra show the slab model that fits to the observations, using a $N_{\mathrm{CO}_2} = 3 \times 10^{15}$ cm$^{-2}$ and T = 700K. In contrary to the 15 $\mu$m feature, the thermo-chemical models, are all significantly weaker than the slab model.}
    \label{fig:CO24micron}
\end{figure*}

Aside from the strong feature at 15 $\mu$m, \ce{CO2} has another strong vibrational band at 4.3 $\mu$m corresponding to the asymmetric stretch, the 00$^0$1(1)--00$^0$0(1) transition. This transition has an upper level energy of 3350 K and high Einstein A coefficient ($>10$ s$^{-1}$). JWSRT-NIRSpec is likely the first instrument that will be able to detect this band towards proto-planetary disks. At an R$\approx$3000 the lines in the bands will not be entirely separated, but the line peaks should be distinguishable \citep[e.g.][]{Bosman2017}. Fig.~\ref{fig:CO24micron} shows the \ce{CO2} model spectra at 4.3 $\mu$m. 
 
As with the 15$\mu$m feature, including water UV-shielding and chemical heating lowers the total flux in the \ce{CO2} band, due to the lower \ce{CO2} column in the disk model. Truncating the \ce{CO2} emitting region has less of an effect on the 4.3$\mu$m feature than it has on the 15$\mu$m feature. This is due to the higher upper level energy of the transition leading to a more centrally weighted emission profile. It is worth noting that in all cases, the predicted 4.3 $\mu$m flux from the thermo-chemical models is below that of the slab model that fits the 15 $\mu$m flux. As with the water 6.5 $\mu$m emission, this is due a sub-thermal excitation of \ce{CO2} in the surface layers of the disk \citep{Bosman2022water}. This effect is especially notable as the critical density for the 4.3$\mu$m feature is $\sim10^{15}$ cm$^{-3}$ \citep{Bosman2017}. As the predicted continuum around 4.3 $\mu$m for this model is 0.45 Jy, a high continuum S/N, $>$150, is required to detect the full, truncated model. 

\end{document}